\documentclass[fleqn,usenatbib]{mnras}

\usepackage[T1]{fontenc}

\DeclareRobustCommand{\VAN}[3]{#2}
\let\VANthebibliography\thebibliography
\def\thebibliography{\DeclareRobustCommand{\VAN}[3]{##3}\VANthebibliography}

\usepackage{graphicx}	
\usepackage{svg}		
\usepackage{amsmath}	
\usepackage{amssymb}	
\usepackage{multirow}   

\usepackage{newtxtext,newtxmath}

\newcommand{\orcid}[1]{\href{https://orcid.org/#1}{\includesvg[width=10pt]{orcid_logo.svg}}}

\title[Deviations in the IMFs of binary star clusters]{Deviations from the universal Initial Mass Function in binary star clusters}

\author[Sunder S.K. Singh-Bal, G. A. Blaylock-Squibbs, R. J. Parker and S. P. Goodwin]{
  Sunder S. K. Singh-Bal$^{1}$, 
  George A. Blaylock-Squibbs$^{1,2}$, 
Richard J. Parker$^{1}$\thanks{Email: r.parker@sheffield.ac.uk}\thanks{Royal Society Dorothy Hodgkin fellow}, and 
Simon P. Goodwin$^{1}$
\\
$^{1}$Astrophysics Research Cluster, School of Mathematical and Physical Sciences, University of Sheffield, Hicks Building, Sheffield S3 7RH, UK \\
$^{2}$Jeremiah Horrocks Institute, University of Central Lancashire, Preston, Lancashire PR1 2HE, UK}

\date{Accepted XXX. Received YYY; in original form ZZZ}

\pubyear{2024}

\begin{document}
\label{firstpage}
\pagerange{\pageref{firstpage}--\pageref{lastpage}}
\maketitle

\begin{abstract}
The stellar mass distribution in star-forming regions, stellar clusters and associations, the Initial Mass Function (IMF), appears to be invariant across different star-forming environments, and is consistent with the IMF observed in the Galactic field. Deviations from the field, or standard, IMF, if genuine, would be considered strong evidence for a different set of physics at play during the formation of stars in the birth region in question. We analyse $N$-body simulations of the evolution of spatially and kinematically substructured star-forming regions to identify the formation of binary star clusters, where two (sub)clusters which form from the same Giant Molecular Cloud orbit a common centre of mass. We then compare the mass distributions of stars in each of the subclusters and compare them to the standard IMF, which we use to draw the stellar masses in the star-forming region from which the binary cluster(s) form. In each binary cluster that forms, the mass distributions of stars in one subcluster deviates from the standard IMF, and drastically so when we apply similar mass resolution limits as for the observed binary clusters. Therefore, if a binary subcluster is observed to have an unusual IMF, this may simply be the result of dynamical evolution, rather than different physical conditions for star formation in these systems. 
\end{abstract}




\section{Introduction}

The majority of stars form in groups with tens, to thousands, of other stars \citep{Lada03,Bressert10}. Some of these groups become long-lived star-clusters \citep{Kruijssen12b}, although most seem to be part of association-like complexes \citep{Wright20,Wright22} which dissolve into the Galactic disc on relatively short ($<$20\,Myr) timescales.

The distribution of stellar masses -- the Initial Mass Function -- appears largely invariant across many different astrophysical environments (see reviews by e.g.\,\,\citealp{Bastian10,Offner14,Hennebelle24}, though see \citealp{Dib10}, \citealp{Dib18c,Dib22,Dib23},  \citealp{Matzner24} and \citealp{Tanvir24} for arguments to the contrary, and \citealp{Guszejnov19} for arguments against universality outside of the Milky Way). \citet{Bastian10} assert  that there is very little difference between the IMFs of bound, dense star clusters, and the IMFs of less dense and unbound stellar associations. We might therefore expect any observed variations in the IMF between star clusters, or a different IMF from that observed in the Galactic field, to indicate a significant deviation from the physics of star formation in the majority of stellar clusters and associations.

A clustered environment in which we would not expect variations in the IMF is in so-called binary clusters \citep{Slesnick02}. These are two (sub) clusters which orbit a common centre of mass (they do not appear to simply be chance projections), and are fairly common in both the Milky Way \citep{Vereshchagin22} and in the LMC \citep[up to 40\,per cent of open clusters may be part of a binary pair,][]{delaFuenteMarcos09}. Because of their close proximity and presumed shared orbit, they are thought to have formed within the same Giant Molecular Cloud \citep{Dieball02,Dalessandro18,Song22} and therefore we would expect the subclusters to have formed at similar (or identical) metallicities \citep{DeSilva15}, and gas densities \citep{Casado23}.

Whilst their formation mechanisms are currently unclear \citep[e.g. a small number may form via capture,][]{Camargo21}, binary clusters can form from the dynamical evolution of a single star-forming region. In some simulations of kinematically substructured star-forming regions, the stars coalesce into two (or more) subclusters \citep{Parker14b,Arnold17,Parker18b,Schoettler19,Darma21,BlaylockSquibbs23}, which appear similar to the observed binary clusters \citep{Arnold17,Darma21}. Indeed, in their simulations, \citet{Darma21} reproduce the frequency of observed binary clusters (20 -- 40\,per cent), and the mass ratios of the subclusters.

In these simulated star-forming regions, the stars are drawn from an initial mass function that is consistent with the Galactic field IMF \citep[e.g.][]{Maschberger13}. However, the subclusters often have a low mass ratio \citep[one of the subclusters contains significantly more stars than the other,][]{Darma21} and the massive stars often all appear to congregate in just one of the subclusters \citep{Parker14b,BlaylockSquibbs23}.

In our previous work we did not analyse the mass functions of the subclusters to look for variations, and deviations from the standard IMF.  If binary clusters do form from the evolution of a single unbound star-forming region \citep[as postulated by][]{Arnold17} with a universal IMF, how often are the IMFs of the subclusters statistically different?

In this work we ask whether a `standard' IMF can be altered due to the dynamical evolution of a star-forming region and the formation of a binary star cluster. We recognise, however, that this does not test whether a non-universal IMF would subsequently evolve to resemble a universal IMF, in either binary or single star clusters \citep[see e.g.][for an example of dynamical evolution of non-standard IMFs in single clusters]{Kouwenhoven14}.

In this paper, we identify binary clusters in similar simulations to those in \citet{Arnold17}, and then compare the stellar mass distributions of the subclusters to look for variations from the IMF from which the stellar masses for the birth star-forming regions were drawn. The paper is organised as follows. In Section~\ref{sec:methods} we describe the set-up and execution of the $N$-body simulations. We present our results in Section~\ref{sec:analysis}, we provide a discussion in Section~\ref{sec:discussion} and we conclude in Section~\ref{sec:conclusions}. 

\section{Methods}
\label{sec:methods}

In this section we describe the set-up of our $N$-body simulations, before describing the alogorithms we use to identify binary star clusters in the simulations. 

\subsection{$N$-body simulations}
\label{sec:simulations}

The simulations contain $N_\star = 1000$ stars drawn from a \citet{Maschberger13} IMF with a probability distribution of the form
\begin{equation}
p(m) \propto \left(\frac{m}{\mu}\right)^{-\alpha}\left(1 + \left(\frac{m}{\mu}\right)^{1 - \alpha}\right)^{-\beta}.
\label{maschberger_imf}
\end{equation}
Here, $\mu = 0.2$\,M$_\odot$ is the characteristic stellar mass, $\alpha = 2.3$ is the \citet{Salpeter55} power-law exponent for higher mass stars, and $\beta = 1.4$ describes the slope of the IMF for low-mass objects \citep*[which also deviates from the log-normal form;][]{Bastian10}. We randomly sample this distribution in the mass range 0.1 -- 50\,M$_\odot$. This results in a total mass for each region between 550 and 650\,M$_\odot$, with the variation simply due to stochastic sampling of this function.

In previous papers \citep[e.g.][]{Parker14b,Arnold17}, we have shown that binary star clusters can form from the dynamical evolution of kinematically substructured star-forming regions. In those simulations, relatively close stars have small velocity dispersions \citep[cf.][]{Larson82}, which enables the long-term survival of substructure in unbound star-forming regions. Substructure in star-forming regions that are bound tends to be erased by dynamical encounters. However, \citet{Parker14b} show that in the absence of kinematic substructure (small velocity dispersions in the spatial substructure), unbound star-forming regions also erase substructure.

We set up our simulations with substructure using the box-fractal method \citep{Goodwin04a,Cartwright04,DaffernPowell20}. The method is described in detail in the aforementioned cited papers, but we describe it briefly again here.

We set up a cube with side length $N_{\rm div} = 2$, which is then divided into $N_{\rm div}$ smaller sub-cubes. A particle is placed at the centre of each sub-cube and the probability of that particle's cube being subdivided is $N_{\rm div}^{D-3}$, where $D$ is the fractal dimension. A low fractal dimension (e.g.\,\,$D = 1.6$), means the probability of a cube maturing is low, which terminates the subdivision and creates a substructured distribution. A high fractal dimension (e.g.\,\,$D = 3.0$) leads to a high probability of a cube maturing and subdividing again, resulting in a much smoother distribution. We adopt a fractal dimension of $D = 1.6$ in our simulations.

The particles at the final generation of subdivision become the stars in the simulation, and are assigned a small amount of positional noise to prevent the fractal having a grid-like appearance. The velocities of the first generation of particles are drawn from a  Gaussian of mean zero, and the subsequent generations in the subdivision inherit this velocity, plus a small random component that decreases through each subdivision. This results in nearby stars having very similar velocities, but the velocities of distant stars can be very different.

Finally, the velocities of the stars are scaled to a virial ratio $\alpha_{\rm vir} = T/|\Omega|$, where $T$ is the total kinetic energy and $|\Omega|$ is the total gravitational potential energy of the stars, respectively. Our simulations are slightly supervirial, where $\alpha_{\rm vir} = 0.9$ ($\alpha_{\rm vir} = 0.5$ is virial equilibrium).

We do not include primordial binary stars in the simulations.

We run twenty versions of the same simulation, identical apart from the random number seed used to initialise the masses, positions and velocities of the stars. We evolve the star-forming regions as pure $N$-body simulations using the 4$^{\rm th}$-order Hermite integrator \texttt{kira} within the \texttt{Starlab} environment \citep{Zwart99,Zwart01}.

As discussed in \citet{Parker14b} and \citet{Arnold17}, the evolution of these supervirial, spatially and kinematically substructured star-forming regions can result in very different morphologies. In addition to our binary clusters, some simulations can remain highly filamentary, or form three or four subclusters. In this work we analyse only the simulations that formed binary clusters.

The simulations are run for 10\,Myr and we check whether there are binary clusters every 0.1\,Myr throughout the simulations.


\subsection{Binary cluster identification}
\label{sec:cluster_identification}

To robustly identify binary clusters, we first check for distinct subclusters in space using DBSCAN \citep{ester_density-based_1996}. DBSCAN groups points 
together that are within a specified search radius of one another, and then 
discards groups that have too few points in them. We use the implementation of DBSCAN in the scikit-learn Python package \citep{pedregosa_scikit-learn_2011}, adopting a search radius of 3\,pc and a minimum subcluster size of $N = 10$.  As our simulated star-forming regions expand from initial radii of 1\,pc, to radii of 10s\,pc, this choice of search radius facilitates the robust indentification of subclusters when they form.

We then use INDICATE \citep{Buckner19,BlaylockSquibbs22} to assess the level of clustering of each individual star assigned to each subclusters. INDICATE works by comparing a distribution of stars to a uniform control grid of the same average density as the distribution of stars. The average distance, $\bar{r}$, to the $n^{\rm th}$ nearest neighbor is calculated for the control grid. For each individual star, we determine how many  stars are within $\bar{r}$ compared to the average; if this number is significantly higher the star is said to be clustered.

Finally, once we have identified stars in each subcluster using DBSCAN and INDICATE, we check that the total energy of each star is negative, meaning that the star is gravitationally bound to the subcluster.

We note that \citet{Darma21} use an alternative method to find binary clusters in their simulations. They use a minimum spanning tree (MST) to link all of the stars via a single path, and then identify groups based on whether a star lies more than a certain MST branch length away from other stars. The stars that are less than a certain branch length from their connecting star in the MST are grouped together. Similarly, \citet{Parker18b} identify groups using the Friends-of-Friends algorithm, which groups stars based on nearest neighbour distances and velocities. If we adopted either method we would likely identify the same binary clusters, but the exact stellar membership may vary.

\section{Results}
\label{sec:analysis}

We analyse a suite of 20 simulations of supervirial (expanding) star-forming regions and visually identify two that subsequently form a distinct binary cluster system, using DBSCAN and INDICATE as outlined in Section~\ref{sec:cluster_identification}. We analyse each of the simulations in the $x-y$ plane, $x-z$ plane, $y-z$ plane, and in 3D.

\begin{figure}
 	\includegraphics[width=\columnwidth]{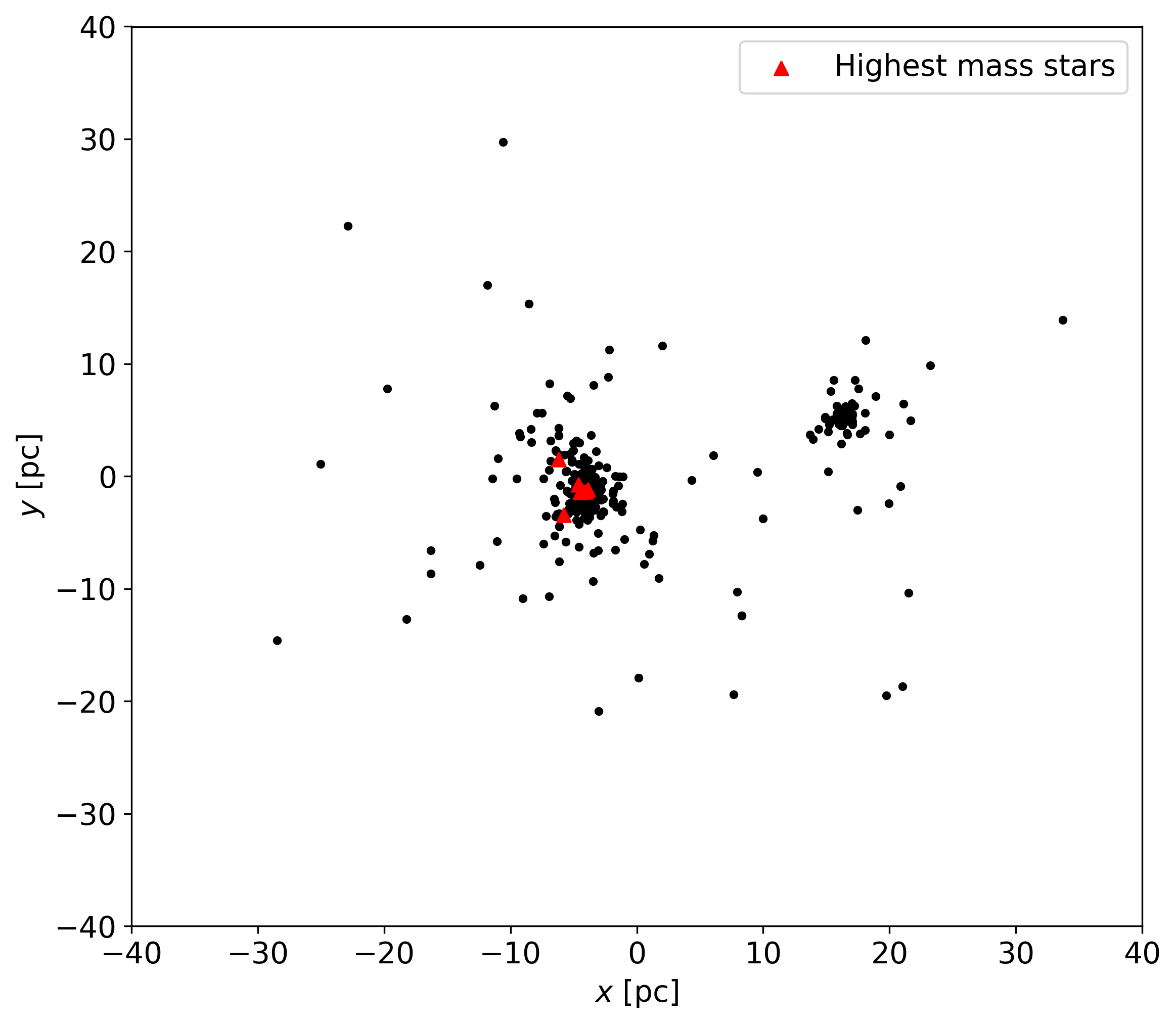}
        \caption{A snapshot after 5\,Myr of evolution of a star-forming region in the first of our two simulations that forms a binary cluster. The ten most massive stars are shown by the red triangles, and all are located in one of the subclusters.}
     \label{fig:highlight}
 \end{figure}

We show a snapshot from one of our two simulations that forms a binary cluster in Fig.~\ref{fig:highlight}. Whilst two distinct subclusters are clearly visible, the ten most massive stars (shown by the red triangles) are all located in the more massive subcluster.

This simulation forms a binary cluster from just after 2\,Myr, which remains until the end of the simulation. The snapshot we show in Fig.~\ref{fig:highlight} is at 5\,Myr.

\begin{figure}
 	\includegraphics[width=\columnwidth]{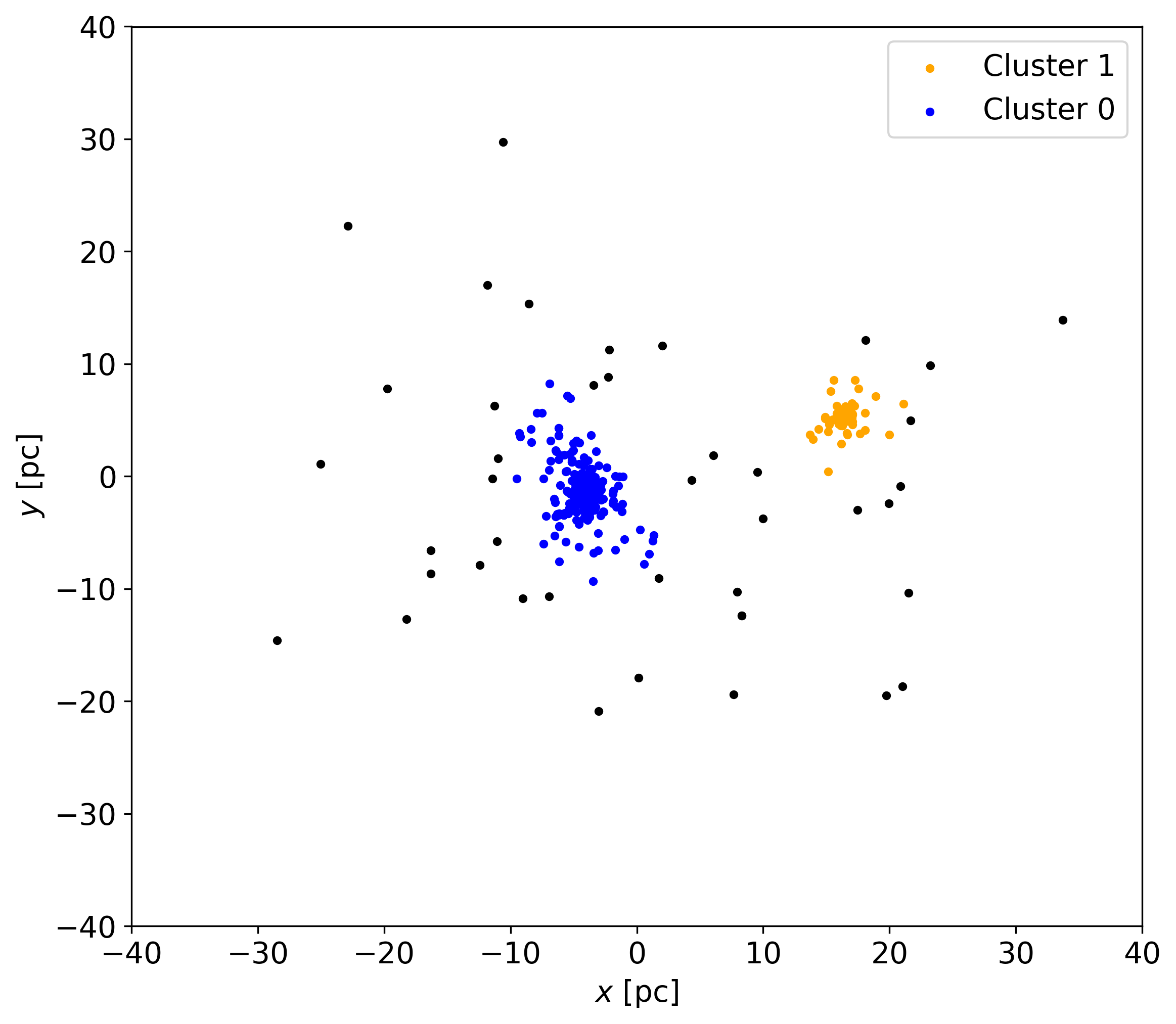}
        \caption{Plot showing an example binary cluster from our simulations at an age of 5\,Myr, with the stars coloured according to the subclusters there are  assigned to with DBSCAN. The black points are stars that are not assigned to either subcluster.}
     \label{fig:dbscan}
 \end{figure}

In Fig.~\ref{fig:dbscan} we show the stars in this simulation (again after 5\,Myr of evolution) identified as members of the two subclusters with DBSCAN and INDICATE. Members of the more massive/populous subcluster (`Cluster~0') are shown in blue  (285 stars with a total mass of 356\,M$_\odot$), and members of the lower mass subcluster (`Cluster~1') are shown in orange (112 stars with a total mass of 85\,M$_\odot$). Stars not assigned to either cluster are shown by the black points. The mass distribution of stars in each of these subclusters are then compared to a standard \citet{Maschberger13} IMF.

Many of the observed binary clusters are located at significant distances from the Sun (kpcs) and so the lower mass limit for stars that can be individually resolved can be as high as 1\,M$_\odot$. We perform our analysis on the entire simulation data, before restricting the sample to stars with masses exceeding 0.2 M\textsubscript{\(\odot\)}, 0.3 M\textsubscript{\(\odot\)} and 0.4 M\textsubscript{\(\odot\)}.

For each subcluster, we perform one-sample KS tests \cite[]{daniel_applied_1990} and one-sample Cram{\'e}r-von Mises tests 
\cite[]{csorgo_exact_1996} to compare the mass distribution of each 
subcluster to the standard Maschberger IMF used to generate the whole population. 

The minimum KS-test $p$-values for each simulation that forms a binary cluster are shown in Table~\ref{table:sim_7} and Table~\ref{table:sim_8}. The columns in each table are the lower mass limit (below which stars are not included in the KS test), and then the KS $p$-value for each subcluster in  various projections. Where the $p$-value falls below 0.1, we reject the null hypothesis that the mass function in the subcluster shares the same underlying parent distribution as the normal \citet{Maschberger13} IMF used to set up the masses in the simulations.

\begin{figure}
 	\includegraphics[width=\columnwidth]{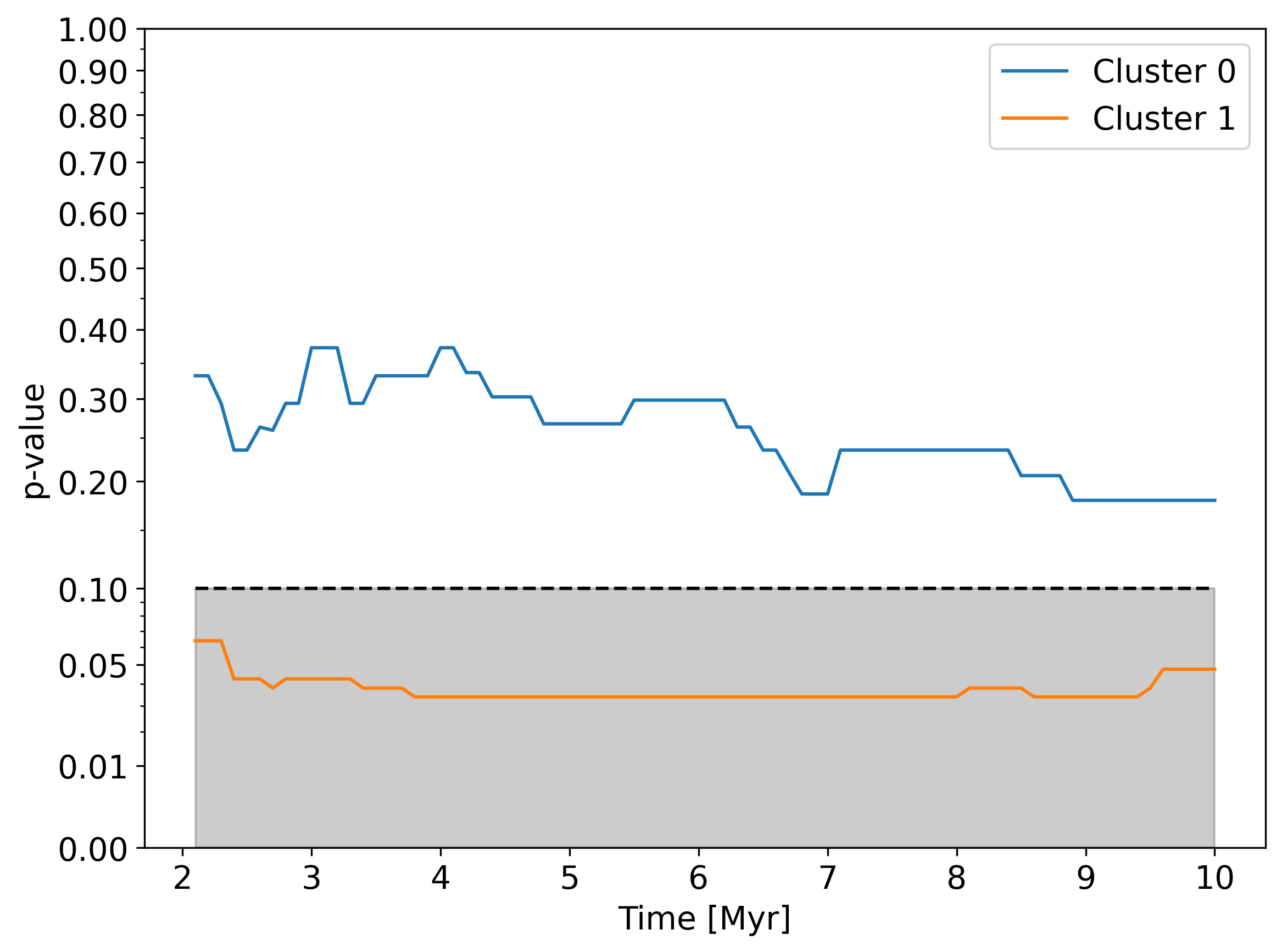}
        \caption{Evolution of the $p$-value for the KS-test between a \citet{Maschberger13} IMF and the mass function in each of the subclusters of the binary cluster that forms after 2.1\,Myr in our first simulation. The blue line corresponds to the more massive Cluster~0 and the orange line corresponds to the less massive Cluster~1. The region where the $p-$value from the KS test is less than 0.1 is shown by the grey shaded area below the dashed line. To mimic observational limitations, stars with masses below 0.3\,M$_\odot$ are not included in the analysis.}
     \label{fig:KS_test}
 \end{figure}

In Fig.~\ref{fig:KS_test}, we visualise the evolution of the $3D$ data from our first simulation (Table~\ref{table:sim_7}) where we impose a minimum mass of 0.3\,M$_\odot$ for our IMF comparisons. Each line represents the $p$-value for each subcluster IMF comparison as the simulation evolves (the star-forming region forms a distinct binary cluster after $\sim$2.1\,Myr). In this simulation, the KS test between the mass function of Subcluster 1 (shown by the orange line) and the standard \citet{Maschberger13} IMF suggests the two mass functions do not share the same underlying parent distribution. 

\begin{table*}
    \caption{Minimum KS-test $p$-values for our first simulation in which a binary cluster forms. Columns are the stellar minimum mass in the IMF comparisons, and then the smallest $p$-values calculated from KS-tests between the mass function of the subclusters and the \citet{Maschberger13} IMF, in different projections. $p$-values below 0.1 are shown in bold font.}
    \label{table:sim_7}
    \begin{tabular}{ ccccccccc } 
        \hline
        Lower mass limit & \multicolumn{2}{c}{$x-y$ plane} & \multicolumn{2}{c}{$x-z$ plane} & \multicolumn{2}{c}{$y-z$ plane} & \multicolumn{2}{c}{$3D$} \\
       (M$_\odot$) & Cluster 0 & Cluster 1 & Cluster 0 & Cluster 1 & Cluster 0 & Cluster 1 & Cluster 0 & Cluster 1 \\
        \hline
        0.1 & 0.59820 & 0.48283          & 0.48249 & 0.44278          & 0.75322 & 0.33696          & 0.38451 & 0.37001 \\ 
        0.2 & 0.38424 & 0.17151          & 0.40682 & 0.14866          & 0.48792 & 0.12405          & 0.40006 & 0.14086 \\ 
        0.3 & 0.17972 & \textbf{0.03390} & 0.20263 & \textbf{0.02776} & 0.17919 & \textbf{0.02699} & 0.16804 & \textbf{0.02206} \\ 
        0.4 & 0.32499 & \textbf{0.00746} & 0.42828 & \textbf{0.00516} & 0.26908 & \textbf{0.00842} & 0.28671 & \textbf{0.00516} \\ 
        \hline
    \end{tabular} 
\end{table*}

\begin{table*}
    \caption{Minimum KS-test $p$-values for our second simulation in which a binary cluster forms. Columns are the stellar minimum mass in the IMF comparisons, and then the smallest $p$-values calculated from KS-tests between the mass function of the subclusters and the \citet{Maschberger13} IMF, in different projections. $p$-values below 0.1 are shown in bold font.}
    \label{table:sim_8}
    \begin{tabular}{ ccccccccc } 
        \hline
        Lower mass limit & \multicolumn{2}{c}{$x-y$ plane} & \multicolumn{2}{c}{$x-z$ plane} & \multicolumn{2}{c}{$y-z$ plane} & \multicolumn{2}{c}{$3D$} \\
       (M$_\odot$) & Cluster 0 & Cluster 1 & Cluster 0 & Cluster 1 & Cluster 0 & Cluster 1 & Cluster 0 & Cluster 1 \\
        \hline
        0.1 & 0.64328 & 0.13556          & 0.72191 & 0.39337 & 0.46959 & 0.37119 & 0.73853 & 0.26136 \\ 
        0.2 & 0.56805 & \textbf{0.04982} & 0.72085 & 0.19499 & 0.81819 & 0.19662 & 0.50805 & 0.18027 \\ 
        0.3 & 0.13998 & 0.24801          & 0.15601 & 0.54855 & 0.14256 & 0.52831 & \textbf{0.09353} & 0.33140 \\ 
        0.4 & 0.94218 & 0.48598          & 0.82477 & 0.43458 & 0.79448 & 0.81981 & 0.93308 & 0.35565 \\ 
        \hline
    \end{tabular} 
\end{table*}

 \begin{figure}
 	\includegraphics[width=\columnwidth,trim={0 0.2cm 0 0},clip]{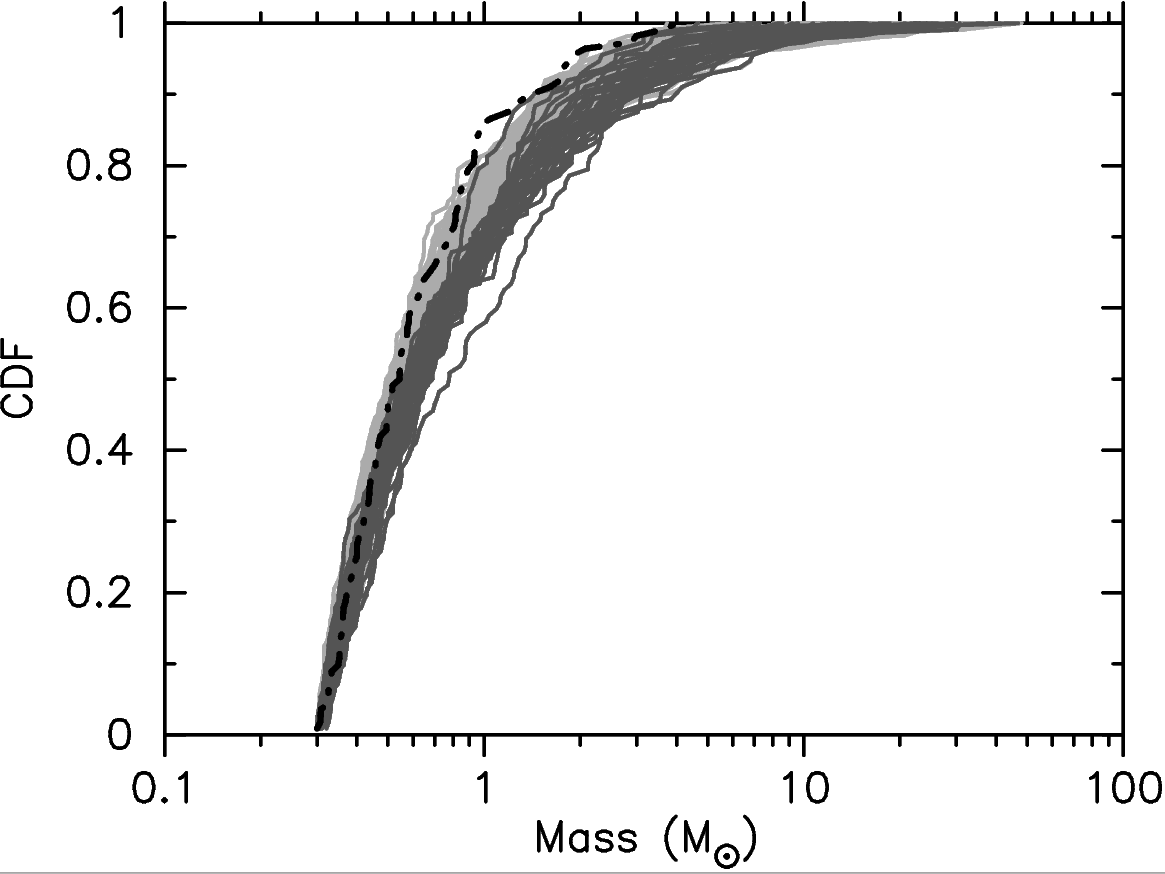}
        \caption{Comparison between the IMF of the stars in Subcluster 1 in our first simulation (Table~\ref{table:sim_7}), shown by the black dot-dash line, and randomly generated IMFs with the same lower-mass cut-off (0.3\,M$_\odot$). Of the 100 randomly generated IMFs (shown by the light grey lines), in 34 a KS test between the randomly generated IMF and the simulation IMF returns a p-value less than 0.1, suggesting we can reject the hypothesis that they share the same underlying parent distribution (these are shown by the darker grey lines). The simulated IMF in Subcluster 1 also appears `bottom heavy' -- it contains no stars more massive than 4\,M$_\odot$, and all of the more massuve stars are situated in Subcluster 0. }
     \label{fig:IMF_comp}
 \end{figure}

The deviation from the input IMF of the simulation in Subcluster~1 is likely due to the massive stars all residing in the other subcluster. This has been routinely observed in similar simulations \cite{Arnold17,Parker18b,Park20}, and is usually simply due to random dynamic motion within the original spatial and kinematic substructure, rather than any differences in the initial conditions. To demonstrate this, we compare the IMF of Subluster~1 to 100 randomly generated IMFs with the same lower-mass cut-off (0.3\,M$_\odot$). This is shown in Fig.~\ref{fig:IMF_comp}, where the dot-dash black line is the IMF from Subcluster~1, and the 100 random realisations are shown by the light grey lines. In 34 of these realisations (shown by the darker grey lines), a KS test between them and the IMF of Subcluster~1 returns a p-value $<$0.1, suggesting we can reject the hypothesis that they share the same underlying parent distribution. This excercise clearly demonstrates the deficiency in high-mass stars in Subcluster~1 in this simulation is responsible for the differences between the two mass functions.

\section{Discussion}
\label{sec:discussion}

Whilst we have shown that binary clusters that form from  a star-forming region with a normal IMF may develop subclusters whose mass functions deviate from the IMF, there are several important caveats to our results.

First, we do not know whether binary clusters do form from the dynamical evolution of an unbound, spatially and kinematically substructured star-forming region \citep{Arnold17}. Whilst some star-forming regions appear to be expanding \citep{Kounkel18}, it is unclear whether they are subtructured to the degree required to produce the binary clusters in our simulations. However, we note that \citet{Darma21} show that binary clusters can also form if the star-forming region is initially in virial equilibrium. 

On a related point, our simulations do not include a background gas potential, and clearly do not react to the removal of any such potential. The simulations are supervirial to begin with, which could mimic early expansion due to gas removal \citep{Tutukov78,Goodwin97a,Baumgardt07,Shukirgaliyev18}; however, if this occurred when substructure was still present it would likely imply that the massive star(s) had formed first, and already started to evolve, before any significant numbers of low-mass stars had formed.

Many observed binary clusters are older systems, and their massive stars may have already left the main sequence, and/or been ejected \citep{Schoettler19,Schoettler20,Farias20}. Therefore, their mass functions may not be \emph{initial} mass functions, and the stellar evolution within binary clusters may act to homogenise disparate mass functions between the subclusters (e.g.\,\,if the massive stars are no longer present).

Observations of relatively nearby Galactic binary clusters do not appear to display any significant variations in the mass functions of the subclusters (e.g. $h$ and $\chi$~Per are conisistent with \citet{Salpeter55} slope mass function, \citealp{Slesnick02}). However, \citet{Bragg05} find that $h$~Per is mass-segregated, but $\chi$~Per is not, and intriguingly, the $h$~Per subcluster is the more massive component (the mass ratio of the subclusters is 0.78, \citealp{Bragg05}). The simulation we present in Fig.~\ref{fig:highlight} is mass-segregated in the most massive of the subclusters, possibly due to the most massive stars all residing in this subcluster and dominating their local potential well \citep{Parker14b,Parker17b}. A similar result is found when simulating clusters close to the Galactic centre \citep{Park20}. \citet{Park20} find that the strong tidal field near the Galactic centre shears apart star-forming regions, but subclusters form in the tidal tails of the sheared regions. These subclusters can be significantly mass-segregated with a top-heavy IMF, or not mass-segregated at all, depending on the (stochastic) dynamical evolution of the star-forming region.  

That there are massive stars in the 12.8\,Myr $h$- and $\chi$-Per system suggests that either the massive stars take significant time to form after low-mass star formation, or that massive stars prolong their lives due to either significant (100\,km\,s$^{-1}$) rotation \citep{Limongi18}, or through mergers \citep{Schneider14}. In our simulations the stars all form at the same time, but this may be a valid assumption in light of these recent developments in our understanding of massive star evolution.

Whilst many studies show that the removal of the gas potential by feedback from massive stars can dominate the dynamical evolution of star-forming regions, analysis of hydrodynamical simulations \citep{Lucas20} suggests that supernovae do not cause the destruction of the star-forming region, but rather the energy from the supernova(e) simply leaks out through the path of least resistance, such as a cavity or low-density part of the gas cloud.

\section{Conclusions}
\label{sec:conclusions}

We analyse $N$-body simulations of the  dynamical evolution of supervirial (unbound) star-forming regions and identify those regions that form binary star clusters -- two subclusters orbiting a common centre of mass. We then compare the mass distributions of stars in the subclusters to a standard Initial Mass Function. Our conclusions are the following:

(i) In a set of twenty simulations, identical apart from the random number seed used to initialise the masses, positions and velocities of stars, two form obvious binary clusters.

(ii) In both simulations, a KS test between the mass distribution of stars in one of the subclusters, and the IMF used as an input to the simulations, returns a low $p-$value, such that we can reject the hypothesis that they share the same underlying parent distribution.

(iii) The apparent deviation from the standard IMF happens less often if we include stars down to the hydrogen burning limit. Whilst most observations of binary clusters are only sensitive to individual stellar  masses of $\sim$0.5\,M$_\odot$, future observations may probe lower masses. We would not expect any deviation from the standard IMF if observations were complete down to 0.1\,M$_\odot$.\\

Our results demonstrate that observed variations in an IMF in binary star clusters can result from the dynamical evolution of a single population of stars with a standard IMF. 

\section*{Acknowledgements} 
SSKSB acknowledges support from the 2023 Sheffield Undergraduate Research Experience (SURE) scheme. GABS acknowledges a University of Sheffield publication scholarship. RJP acknowledges support from the Royal Society in the form of a Dorothy Hodgkin Fellowship. We thank the anonymous referee for a helpful report. 
For the purpose of open access, the authors have applied a Creative Commons Attribution (CC BY) license to any Author Accepted manuscript version arising.


\section*{Data Availability}
The data underlying this article will be shared on reasonable request to the corresponding author.





\bibliographystyle{mnras}
\bibliography{general_ref} 




\label{lastpage}
\end{document}